# Seeking Black Lining In Cloud

**Shuchi Sethi**
Department of Computer Science,
Jamia Millia Islamia,
New Delhi, India
Email Id: shuchi.sethi@yahoo.com

**Kashish Ara Shakil**
Department of Computer Science,
Jamia Millia Islamia,
New Delhi, India
Email Id: shakilkashish@yahoo.co.in

**Mansaf Alam**
Department of Computer Science
Jamia Millia Islamia,
New Delhi, India
Email Id: malam2@jmi.ac.in

*Abstract – This work is focused on attacks on confidentiality that require time synchronization. This manuscript proposes a detection framework for covert channel perspective in cloud security. This problem is interpreted as a binary classification problem and the algorithm proposed is based on certain features that emerged after data analysis of Google cluster trace that forms base for analyzing attack free data. This approach can be generalized to study the flow of other systems and fault detection. The detection framework proposed does not make assumptions pertaining to data distribution as a whole making it suitable to meet cloud dynamism.*

## I. INTRODUCTION

Data loss and theft are both major lurking threats to the cloud and most of the times the way a threat is handled can exacerbate the other by exposing multiple copies of data to risk. Data breach is usually a result of intrusive action that went unnoticed. Present state of the art(about data handling in cloud refer to [7][8]) does guarantee a minimal protection from theft by preventing and detecting intrusion. But any new attack which has distribution similar to the regular events is hard to be noticed. Having said that certain unsupervised techniques may be able to capture or log such event pattern which is not usual. Covert channel attack are such attacks which not only leak sensitive data but remain hidden from detection, situation being amplified by the fact that cloud hosts multiple tenants which can be adversary of one another. Cloud thus provides many avenues of data leak where hypervisor and VM security itself is questionable. In this manuscript we discuss covert channel detection, which in itself is a uphill task given the current state of technology relies on machine learning for anomaly detection. Besides each attack have its own novelty and an ects di erent parameters of the system. This is unsurprisingly true for covert channel attacks as well. Some channels exploit cache, disk, bus, memory and many more shared channels can be exploited for the same. Certain applications like online games can be used for collusion by choosing different move other than the best one [3]. Covert channels have been studied for a long time and found to be non preventable but still may be detectable. No infallible method to detect exists so far, fact brought out by literature review [1],[4].

The method that we propose is based on statistical characteristics of the data. During the analysis of Google cluster trace data based on time of resource request, intervals and access latency parameters. Data collected from our cloud testbed after simulating covert channel attacks was also analysed and there were some interesting observations which are discussed in next section. These observations formed the base of our detection framework. The contributions of this paper are: It perceives the issue of covert channel detection in cloud as a problem of binary classification. It gives certain new observations to enable classification of normal and attack data. A new variable is proposed that potentially serves as an initial threshold for classification framework. A framework for covert channel detection is proposed in cloud and lastly we validate our results by comparing with previous approach in literature. Last section provides empirical evidence [6].

## II. COVERT CHANNEL HIDE AND SEEK

We make an assumption that if an event is not part of an attack then it will be i.i.d that is a random variable will have same probability distribution as others and are mutually independent. Let us have a time series of requests that were made for shared resources, if the requests are timed to create a covert channel then it will not be independent of each other, moreover an IID sequence(independent and identically distributed) does not imply the probabilities for all elements of the event space must be the same. Thus we believe that any time based attack will have time dependency in sequencing the events. Thus we filter out all such sequences.

Algorithm:
1) Collect data of time of request for shared resource: (This can be done by placing a hook code in hypervisor)
2) In the time series data we calculate will be of the form   $x2 = x1 + t$ :

3) Here both x1 and t are random variables; so by

theorem1 x2 is a random variable. Thus if x2 is iid sequence, then data is normal signifying no attack:

4) Else forward it to classification sub algorithm:

In step three we try to filter input data so as to take only the set of data which has any probability of attack. This reduces the time complexity of classification and accuracy is gained. The data will be checked for zero auto correlation on time, signaling its iid property. There are two components of the proposed framework. One is o ine processing (step 2-3) and the other one is active processing (step 4). After obtaining the time series, we preprocess the data to avoid irrelevant results. We normalize the data to speed up the active processing and we obtain first order di erence of data to enable pattern availability. Though the order will depend on the source and environment from which data is being obtained, we were able to obtain the pattern with first order.

Dynamic systems have certain characteristics which we exploit in this manuscript. We obtain time series data which signifies request time and interval of request. Thus $f(x) = x + x$. It would be a good idea to specify a theorem pertaining to random variables. This theorem gives an important result that sum of 2 random variables will be random. Theorem1 : Let X and Y be 2 independent random variables with density functions $fx(x)$ and $fy(y)$ defined for all x. Then the sum $z = x + y$ is a random variable with density function $fz(z)$ where fz is convolution of fx and fy.

## III. DETERMINING CLASSIFYING VARIABLES-THE DECISION RULE

Research begins with identification of distinguishing variables. Various measurable criterias for Inter VM tra c filtering can be: Delay times (inter request time), shape of tra c , individual request sizes request sizes , duration of resource usage, request regularity etc [2]. For time based attacks first natural choice would be to analyze the time operations and intervals of time of resource request. The charts below give a brief idea of the o ine process of variable identification for detection. The first order di erence gives vivid idea of the behavior di erence of time intervals of requesting resource. Following observations can be made: One is the intervals are regular in attack while random element in normal operations data. The source of normal request data is Google cloud trace[5]. The second observation is that normal requests come in bursts and behavior is spiky as contrast to attack behavior. [Figure: 1]

Also for in-depth study of the parameters, Euclidean distance between 2 points is calculated and plotted as shown in [Figure: 2]. this makes the behavior more vivid by another factor. This observation leads to an important conclusion that we cannot apply density based algorithms for detection as they assume attacks are scarce. One important goal for any decision rule is the property of generalization, which is an important property in the field of learning. For this purpose it is required to have an approximation function representing the distribution of the variable. To enable generalization we do not pre-assume any distribution.

## IV. PROPOSED HYPERDIMENSION DETECTION

This section discusses only the core classification mechanism of classifying data. As it is very complex to find out the distribution of data in cloud, we use non parametric approach. Also because output of a dynamical system depends on all previous inputs, it is very complex to use an unsupervised method for the purpose. Non parametric approach does not assume that the structure of a model is fixed and it may grow to accommodate the complexity of the data. Individual variables are typically assumed to belong to parametric distributions, and assumptions about the types of connections among variables are also made. Thus we look at the distribution structure of the resource request time variable [Figure: 3, 4].

The approximation function uses method of least squares provides good approximation for the variable. The [figure: 5] gives Classification Sub algorithm.

4.1. Implementation

Step 1:
By fitting the data we obtain a row vector with certain 0 coefficients.

$$x^6+10x^5+67x^4+271x^3+536x+908 \quad (1)$$

$$2x^7 +30x^6+361x^5+2662x^4+1592x^3+8264x^2+4135x+6626 \quad (2)$$

Step 2:
After obtaining mathematical equations that is define normal and attack behavior based on resource request time. It is important to define threshold values that can be used for comparison. Here normal behavior so obtained will be treated as our threshold.

Distance vector is calculated by subtracting eqn1 from eqn2.

$$2x^7+31x^6+371x^5+2729x^4+1863x^3+8836x^2+4671x+5718 \quad (3)$$

Important Observation:
One point worth mention is that the equation figure attackvector [Figure: 6]. This validates Another observation about attack data is that obtained for attack behavior is a monotonic function as represented in our results as the same observations were made by other researchers. high autocorrelation(r) was observed. Values of r were close to 1.

## V. EVALUATION AND EMPERICAL VALIDATION

Previous approaches that were successfully used in traditional setup for covert channel detection like standard deviation, entropy based approaches were experimented but those did not go well in cloud based attacks in our ex-

Seeking Black lining In cloud

periments. The probable reason for that could be that in cloud so many processes are running at same time and optimization algorithms try to use available resources to the best, leading to less chances of accurate entropy measure-ment for a single process along with smarter attack algorithms which tend to blend in resource requests by neutralizing entropy.

We tested our algorithm on a separate dataset and applied our algorithm. The results are shown in figures below. Also to check monotonicity, we obtained a function using curvefitting. After performing this step there were 0 false negatives while around 20 percent false positives. To bring this down next will check the data for monotonicity.

The figure [ 9] contains two parts first one contains a monotonic function signifying attack while in the second part graph dip signifies normal data behave. The statistical profile of the algorithm performance is given in the table below [table: 10].

### VI. CONCLUSION

Though it is not easy to proof the cloud of all security breaches but as current research is focused on bringing thefts under acceptable limits, authors plan to contribute by taking detection of covert channel further to self-correction mechanism so that instead of logging the system should be able to restore and recover from the attack. Besides as a step further towards generalization, in near future authors plan to propose a framework for anomaly detection in cloud.

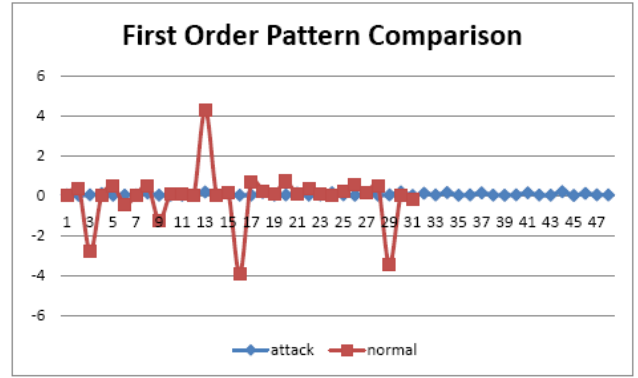

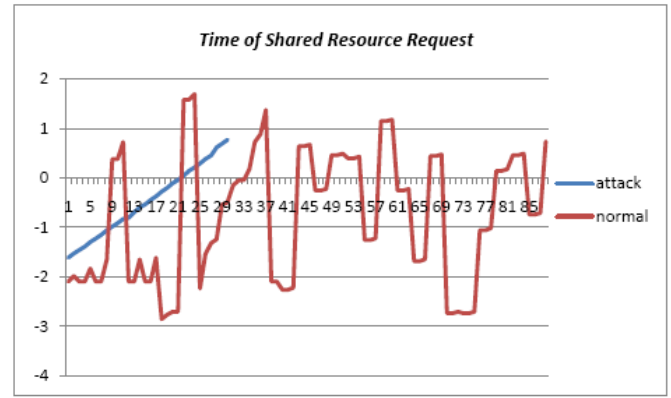

Figure 1: Parameter

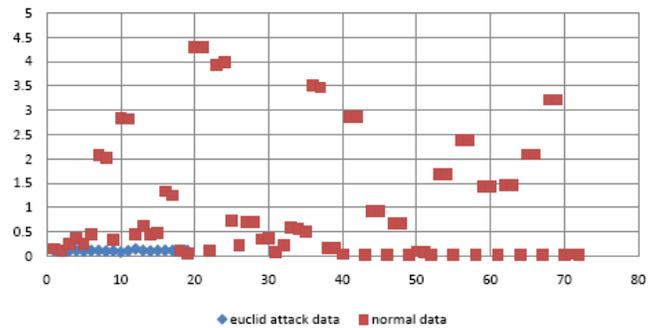

Figure 2: Euclidean Distance

Seeking Black lining In cloud

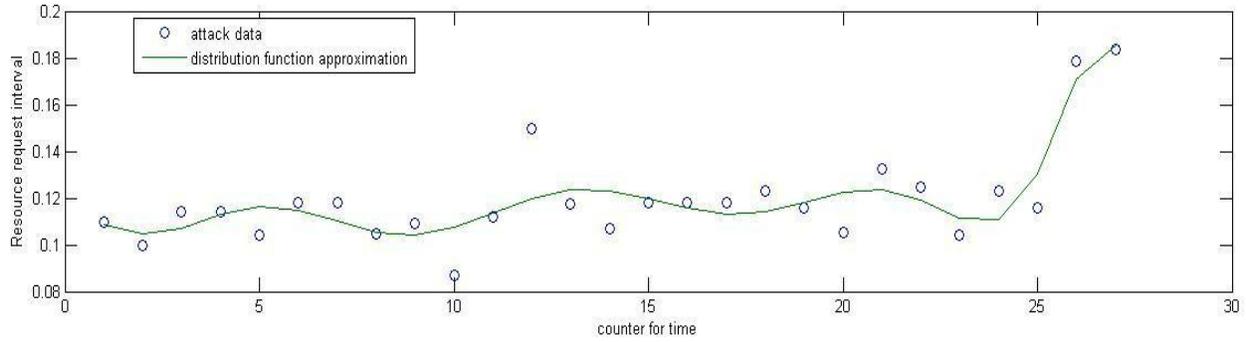

Figure3:Curvefitting

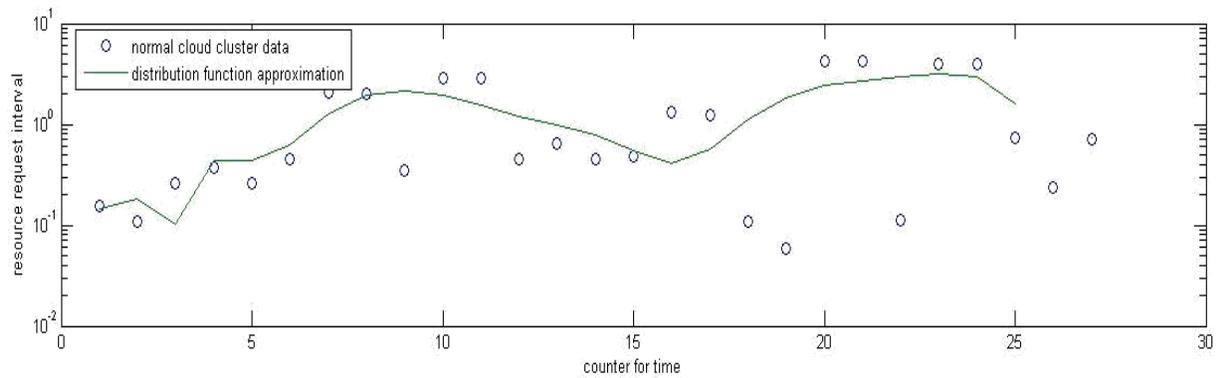

Figure4:Curvefitting

- For each individual value V in forwarded datadet D:
  - Calculate Euclidean distance.
  - Obtain first order series of request time.
- Obtain approximation function for the series(f).
- Calculate autocorrelation(r) and move to next step;
- If r >=0.6

  (this threshold will ensure there are no false negatives) * Move to next step
   Else Print("Take larger interval,not likely to be an attack")

- If (f is a differentiable continuous function in large interval (a,b)) Calculate:
  * if f'(x) > 0 for x in (a,b) then f is increasing
  * if f'(x) < 0 for x in (a,b) then f is decreasing Else Print("Not likely to be an attack")

Figure 5: Classification algorithm

Seeking Black lining In cloud

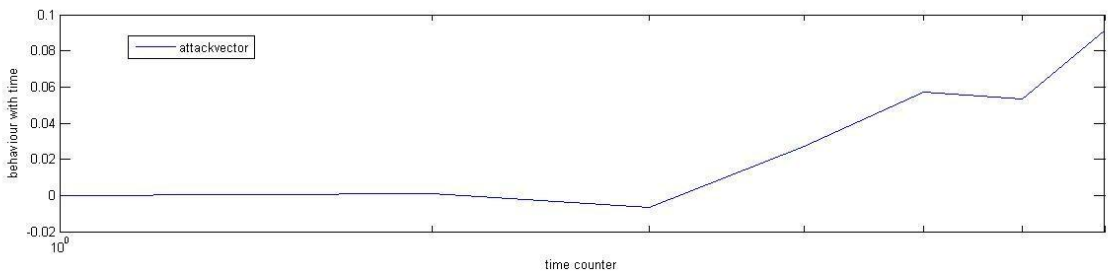

Figure 6 : validation

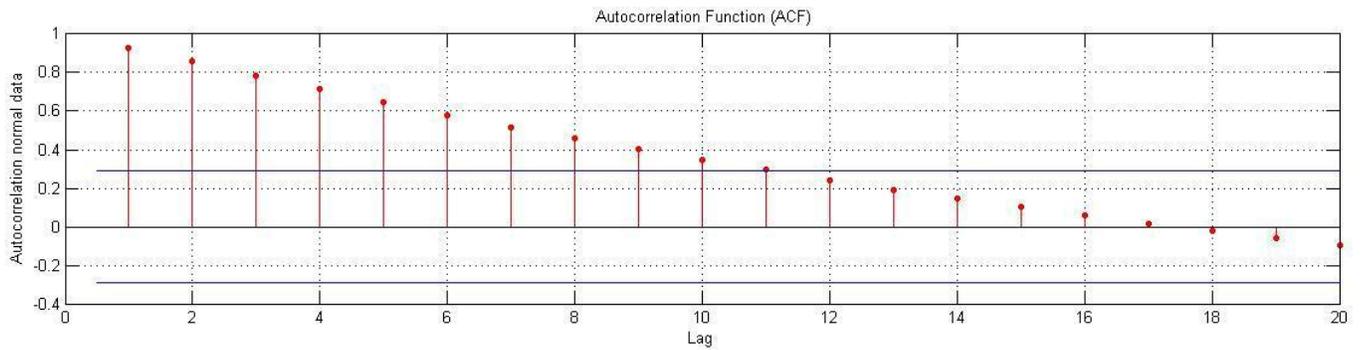

Figure 7: Autocorrelation for validation

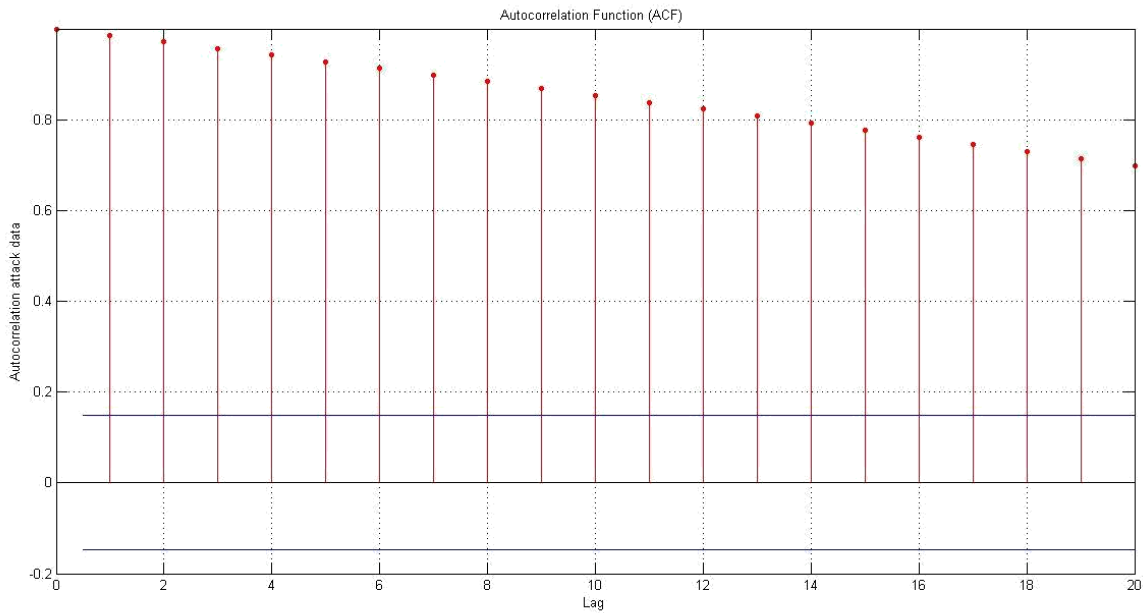

Figure 8: Autocorrelation for validation

Seeking Black lining In cloud

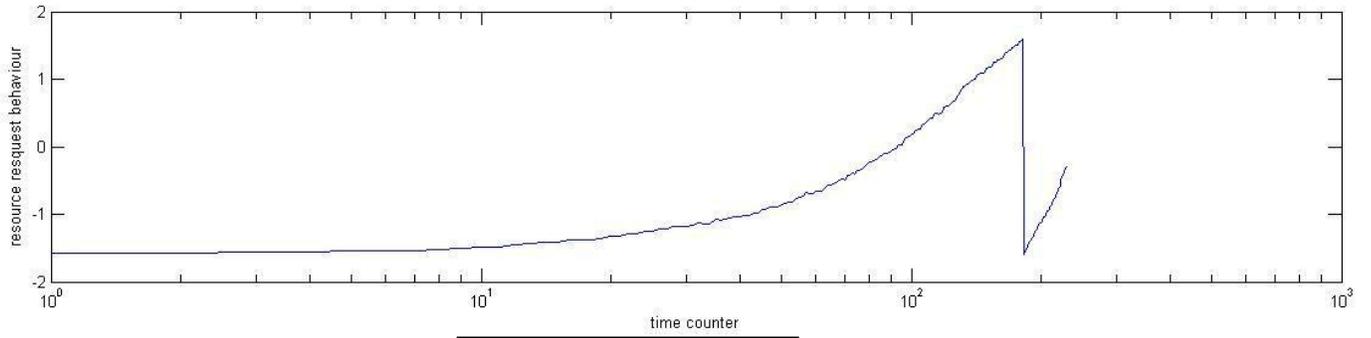

Figure 9 : Function Behaviour

| Channels | False Positives(C2 detector) | False Positives(Detection Engine) | False Negatives(c2 detector) | False Negatives(Detection Engine) |
|---|---|---|---|---|
| CPU load based | 4.46% | 2.28% | Nil | Nil |
| Memory Bus attack | --- | 3% | Nil | Nil |
| Cache Based attack | 1.89 | 1% | Nil | Nil |

Figure 10 : Comparison